\documentclass[reprint,
bibnotes,
amsmath,amssymb,
aps,
]{revtex4-1}
\usepackage{caption}
\usepackage{subcaption}
\usepackage{graphicx}% Include figure files
\usepackage{dcolumn}% Align table columns on decimal point
\usepackage{bm}% bold math
\begin{document}

%\preprint{APS/123-QED}

\title{Magnetic ratchet effect in phosphorene}% Force line breaks with \\

\author{Narjes Kheirabadi }%\email{n.kheirabadi@Lancaster.ac.uk}
\affiliation{Physics Department, Alzahra University, Vanak, Tehran 1993893973, Iran}%
%\affiliation{$^2$ School of Physics, Institute for Research in Fundamental Sciences (IPM), P.O.Box:19395-5531, Tehran, Iran}%


\begin{abstract}
Abstract: The magnetic ratchet effect has been studied in phosphorene by the use of the Boltzmann kinetic equation that is a semi-classical approach. The Hamiltonian of phosphorene in a steady parallel magnetic field is derived using the tight--binding model. We consider the effect of the magnetic field on non--linear dynamics in the presence of an ac laser field and spatial inversion asymmetry. We have shown that for anisotropic 2D materials and phosphorene, the ratchet current has the response to three different light polarizations: linearly polarized light, circularly polarized light, and unpolarized light.        
\end{abstract}

\maketitle
\section{\label{sec:level1}Introduction}
In a ratchet machine, while a pawl moves upward and downward, the wheel rotates in one direction. While the magnetic ratchet effect is an effect accordingly a dc current will be produced in a semiconductor where it is under an alternating electric field of laser radiation and a steady magnetic field. Ratchet effects induced by the in-plane magnetic field were previously studied \cite{tarasenko2008electron, budkin2016ratchet}. It has been experimentally observed in graphene, where the symmetry is broken by adatoms \cite{taranature} or superlattice \cite{PhysRevB.93.075422} and Si-MOSFET \cite{simosfet}. It has also been theoretically predicted for gated bilayer graphene \cite{edratchet, kheirabadi2018cyclotron} and quantum well \cite{direct}.
However, it is unclear how this effect would appear in an anisotropic material. In this paper, we solve this issue by the study of the magnetic ratchet effect in phosphorene, a monolayer of black phosphorus. 

In phosphorene, each phosphorus atom is covalently bonded with three beside phosphorus atoms. Hence, each $p$ orbital retains a lone pair of electrons. Because of the
$sp^3$ hybridization, phosphorene does not form an atomically flat sheet like graphene. This property generates an intrinsic in-plane anisotropy that results in a specific angle-dependent conductivity \cite{viti2015black}. The puckered structure of phosphorene deduces to a strong anisotropy in electric conducting and it is important to have novel devices with anisotropic properties \cite{sun2016optical}. Phosphorene has also been fabricated and manipulated in the lab \cite{lee2019fabrication, akhtar2017recent}. The most remarkable properties of phosphorene are high carrier mobility, high optical and UV absorption, strong in-plane anisotropy, showing a direct bandgap, and other attractive properties, which are of particular interest for optoelectronic applications \cite{akhtar2017recent, carvalho2016phosphorene}. 

Here, we have considered the effect of an ac laser field and an in--plane steady magnetic field on the induced second--order dc current in phosphorene. We have considered the effect of the spatial asymmetry caused by the disorder and an external gate on the ratchet current. 
In this article, we will show that under asymmetric spatial disorder, phosphorene produces a dc current that includes the responses to linearly polarized light, circularly polarized light, and unpolarized light because the momentum relaxation time depends on energy. This effect is observable in monolayer graphene due to the linear dispersion of carriers, as well \cite{taranature}. However, in 2D materials with parabolic dispersion, the circular ratchet effect and the unpolarized ratchet current under an in-plane magnetic field is possible only if the momentum relaxation time depends on energy \cite{falko1989rectifying,direct,edratchet}. As an example, for bilayer graphene where Coulomb impurities act like short--range scatters, it is shown that only the linear magnetic ratchet current is possible \cite{edratchet}. 
Additionally, the established circular polarized light detectors rely on chiral organic semiconductors and metal metamaterials \cite{chen2019circularly}, however, based on this study we predict that inorganic phosphorene could act as a polarized light detector. This point is also valuable to develop a polarization detection and measurement based on anisotropic 2D materials. 
\section{\label{sec:level2}Hamiltonian}
\begin{figure}[h!]
   \centering   
   \includegraphics[scale=0.40]{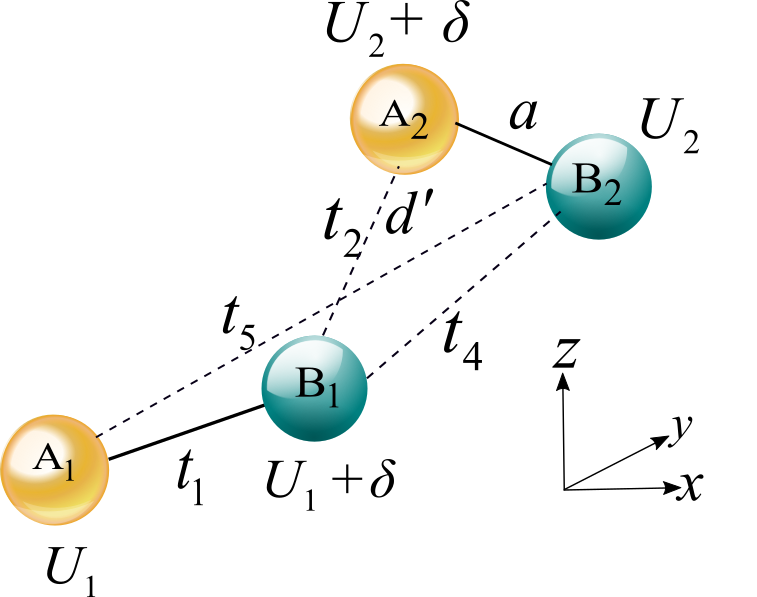}
   \caption[The unit cell of phosphorene]{The side view of four atoms in the unit cell of phosphorene. $A_1$ and $B_1$ atoms on the lower layer, and $A_2$ and $B_2$ on the upper layer have been depicted. The intralayer distance between atoms in one unit cell is $a$, and $d^\prime$ is the distance between $B_1$ and $A_2$ atoms in different layers. Straight lines indicate intralayer coupling $t_1$. Interlayer couplings $t_2$, $t_4$, $t_5$ are depicted by dash lines. Parameters $U_1$, $U_2$, $\delta$ indicate different on--site energies, as described in the main text. Note that $x$ is armchair and $y$ is zigzag direction.} 
    \label{unitcell}
\end{figure}

The unit cell of phosphorene is depicted in Fig.~\ref{unitcell} with armchair (zigzag) edge structure at the $x (y)$ direction. According to this figure, there are four atoms in the unit cell, two atoms on the lower layer ($A_1$ and $B_1$), and two atoms on the upper layer ($A_2$ and $B_2$).
Intralayer coupling $t_1$, interlayer couplings $t_2$, $t_4$ and $t_5$, different on-site energies ($U_1$ and $U_2$) and $\delta$, interlayer potential asymmetry, are depicted in Fig.~\ref{unitcell}. It is worthy to mention that the on-site energy difference $\delta$ between the positions $A_1$ and $B_1$ ($A_2$ and $B_2$) is introduced and determined by infrared spectroscopy \cite{novo}. In addition, the interalayer hopping parameter $t_3$ is the transfer energy of $B_1$ atom of one unit cell with $A_1$ atom of the beside unit cell and it is not depicted in the Fig.~\ref{unitcell}. Furthermore, the intralayer distance between atoms in one unit cell is $a$, and for $d$ as interlayer distance, $d^{\prime}$ is the distance between $B_1$ and $A_2$ atoms. We also assume that two angles $\alpha$ and $\beta$ are $\alpha=\widehat{A_1B_1A_1}=\widehat{B_2A_2B_2}= 96.5 ^ \circ$ and $\beta=\widehat{A_2B_1A_1}-90^\circ=108.04^{\circ}-90^\circ=18.04 ^ \circ$. While, the upper layer is located at $d/2$, the lower layer is located at $-d/2$ where $d = 2.13 \AA$ \cite{chaves2017theoretical, pereira2015landau}. Finally, we assume that phosphorene is under the effect of an in--plane steady magnetic field, $\textbf{B}=(B_x,B_y,0)$, where its vector potential is $\textbf{A}=z(B_y,-B_x,0)$ chosen to preserve translation symmetry in the phosphorene plane; $z$ is the Cartesian coordinate perpendicular to the phosphorene plane.

Because there are four atoms in the unit cell of phosphorene, phosphorene tight--binding Hamiltonian is a $4 \times 4$ matrix, and phosphorene has two conduction bands and two valence bands. To write the tight--binding Hamiltonian of phosphorene in a parallel magnetic field, we use the Peierls substitution. For instance, to determine the Hamiltonian element for a process of hopping between the in--plane A and B sublattices, $H_{A B}$, we have determined the following summation over B sites at the position $\textbf{R}_{B_j}$
\begin{eqnarray}\label{ahbov}
H _ {A B} = t_1 \sum_{j = 1}^{3} \exp \left( i \textbf{K}\cdot(\textbf{R}_{B_j} - \textbf{R}_{A}) - \frac{ie}{\hbar}\int_{R_{B_j}}^{R_{A}}\textbf{A}.d \textbf{l}\right).\nonumber
\end{eqnarray}
Here, $\textbf{K}=\textbf{p}/\hbar$ is the electron wave vector, $-e$ is the charge of electron and $d \textbf{l}$ is the length differential. Consequently, we can show that the Hamiltonian of phosphorene in the steady magnetic field and in the basis of $(A_1,B_1,A_2,B_2)^{T}$ is
\begin{eqnarray}\label{HBP1}
H=\begin{pmatrix}
U_1 & f_1+f_3  & f_4  & f_2+f_5 \\ 
{f_1}^*+{f_3}^* & U_1 + \delta  & {f_2}^*+{f_5}^* & f_4\\ 
{f_4}^* & f_2+f_5 & U_2 + \delta  & f_1^{\prime}+f_3^{\prime} \\ 
{f_2}^*+{f_5}^* & {f_4}^*& {f_1^{\prime}}^*+{f_3^{\prime}}^*  &U_2 
\end{pmatrix}.
\end{eqnarray}
We assume that $\textbf{b}=e d \textbf{B} / 2 $, in--plane momentum $\mathbf{p}=(p_x,p_y,0)$ and $a_x$ and $a_y$ are the length of the unit cell into the $x$ and $y$ directions, respectively. For the lower layer, we have
\begin{eqnarray}\label{f1}
f_1= 2 t_1 \cos{\frac{a_y(p_y+b_x)}{2 \hbar}}\exp{\big[\frac{i(p_x-b_y )}{\hbar}a \cos{\frac{\alpha}{2}}}\big],
\end{eqnarray}
\begin{eqnarray}\label{f3}
f_3 && = 2 t_3 \cos{\frac{a_y(p_y+b_x)}{2 \hbar}}\nonumber\\
&&\quad \times \exp{\big[-\frac{i(p_x-b_y )}{\hbar}(2 d ^ \prime \sin \beta + a \cos{\frac{\alpha}{2})}}\big],
\end{eqnarray}
and for the upper layer, we have 
\begin{eqnarray}\label{f1prime}
f_1^\prime = 2 t_1 \cos{\frac{a_y(p_y-b_x)}{2 \hbar}}\exp{\big[\frac{i(p_x+b_y )}{\hbar}a \cos{\frac{\alpha}{2}}}\big],
\end{eqnarray}
\begin{eqnarray}\label{f3prime}
f_3^\prime && = 2 t_3 \cos{\frac{a_y(p_y-b_x)}{2 \hbar}}\nonumber\\
&&\times \exp{\big[-\frac{i(p_x+b_y )}{\hbar}(2 d ^ \prime \sin \beta + a \cos{\frac{\alpha}{2})}}\big].
\end{eqnarray}
Furthermore, we have
\begin{eqnarray}\label{f2}
f_2= t_2 \exp[-\frac{i}{\hbar}p_x d^\prime \sin\beta],
\end{eqnarray}
\begin{eqnarray}
f_4= 4 t_4 \cos[\frac{p_x}{\hbar}(d^\prime \sin\beta+a \cos\frac{\alpha}{2})] \cos[\frac{p_y}{\hbar}a\sin\frac{\alpha}{2}],
\end{eqnarray}
\begin{eqnarray}\label{f5}
f_5= t_5 \exp[i \frac{p_x}{\hbar}(a_x-d^\prime \sin\beta)].
\end{eqnarray}
Additionally, it is important to work in the low--energy regime. To do so, we make a Taylor expansion of $f_i$ functions in the vicinity of the $\Gamma$ point. 
Consequently, we can assume that $\cos x = 1 - x^2 / 2$ and $\exp x = 1 + x + x^2/2$. In addition, we neglect those terms that are quadratic or higher in the magnetic field.
\section{Ratchet Current in a two dimensional material}
According to the perturbation theory, the magnetic dependent valence band is
\begin{eqnarray}\label{perteigen}
| 0 \rangle^{p}=|0\rangle+\frac{\langle1| V |0\rangle}{\epsilon _1 - \epsilon _0}{|1\rangle},
\end{eqnarray}
where $ |0\rangle$ and ${ |0 \rangle}^ p =  | \mathbf{p}\rangle$ are valence band and perturbed valence band eigenstates, respectively. In this equation, $|1\rangle$ is the conduction band eigenstate, and $V$ is that part of the Hamiltonian which includes the magnetic field. Additionally, $\epsilon_1$ and $\epsilon_0$ are the conduction band and the valence band energies, respectively. By the same method, the perturbed  conduction band is derived, as well.
 
We assume that the two dimensional material, phosphorene, is under an in--plane ac electric field with the angular frequency $\omega$. It means that $\mathbf{E}_{\parallel}(t)=\mathbf{E}_{\parallel}exp(-i \omega t)+\mathbf{E}_{\parallel}^* exp(i \omega t)$, where $\mathbf{E}_{\parallel}=(E_x,E_y)$ and $\mathbf{E}_{\parallel}^*=(E_x^*,E_y^*)$. This in--plane radiation changes the electron distribution function, so that the electron distribution function is dependent on the momentum $\mathbf{p}$ and $t$ time; $f(\mathbf{p},t)$. For homogeneous materials, we use Boltzmann kinetic equation, hence, the results of this paper are valid in the semi-classical regime, $\hbar \omega \ll \epsilon _f$; $\epsilon _f$ is the Fermi level \cite{glazov2014high}. Assuming $\mathbf {v}\cdot \partial f / \partial \mathbf{r}=0$, $\mathbf{v}$ is the electron velocity, we have 
\begin{eqnarray}\label{bke}
- e E_{\parallel} \cdot \nabla_p f ( \mathbf {p} , t)+\frac{\partial  f ( \mathbf {p} , t)}{\partial t}=S \{f\}. 
\end{eqnarray}
Collision integral $S \{f\}$ is
\begin{eqnarray}
S \{f \}=\sum_{\mathbf{p^{\prime}}}[W_{\mathbf{pp^{\prime}}}f(\mathbf{p^{\prime}},t)-W_{\mathbf{p^{\prime}p}}f(\mathbf{p},t)].\nonumber
\end{eqnarray}
For a perturbed electron gas, the scattering rate is
\begin{eqnarray}\label{sr}
W_{\mathbf{p^{\prime}p}}=W_{\mathbf{p^{\prime}p}}^{(0)}+ \delta W_{\mathbf{p^{\prime}p}},
\end{eqnarray}
where $W_{\mathbf{p^{\prime}p}}^{(0)}$ is the rate of the electron scattering between unperturbed states, and $\delta W_{\mathbf{p^{\prime}p}}$ is the change of the scattering rate because of the perturbation. 

Additionally, according to the Fermi's golden rule, the transition rate between $\textbf{p}$ and $\mathbf{p^{\prime}}$ states under a scattering potential, $\delta H$, is
\begin{eqnarray}\label{formulsp}
W_{\mathbf{p^{\prime}p}}=\frac{2\pi}{\hbar}\left | \langle \mathbf{p^{\prime}}\left | \delta H \right | \textbf{p} \rangle \right |^2 \delta(\epsilon_\mathbf{p}-\epsilon_\mathbf{p^{\prime}}). 
\end{eqnarray}
where angular brackets indicate an average over impurity positions. Considering static impurities, we can write the following equation for $\delta H$
\begin{eqnarray}\label{deltah}
\delta H=\sum_{j=1}^{N_{imp}}\hat{Y}u(\mathbf{r}-\mathbf{R_j}),
\end{eqnarray}
where $N_{imp}$ is the number of impurities, $u(\mathbf{r}-\mathbf{R_j})$ describes
the spatial dependence of the impurity potential, and $\hat{Y}$ is a dimensionless matrix describing the structure.
We also neglect the interference between different impurities, and we use the Fourier transform of the impurity potential
\begin{eqnarray}
\widetilde{u}(\mathbf{q})=\int d^2 r u(\mathbf{r})e^{-i \mathbf {q}.\mathbf{r}/ \hbar}.\nonumber 
\end{eqnarray}
In the scattering rate, we perform a harmonic expansion of the impurity potential as described in the following equation
\begin{eqnarray}
\mid \tilde {u}(\mathbf{p^{\prime}-p}) \mid ^2 =\sum_{m^{\prime}}\nu_{m^{\prime}}e^{im^{\prime}(\phi^{\prime}-\phi)},\nonumber
\end{eqnarray}
where $\phi$ is the polar angle of momentum and $\nu_{-m} = \nu_m$ because it is an even function of
$(\phi^{\prime}-\phi)$. 

To determine the current by the Boltzmann kinetic equation (Eq.~\ref{bke}), we consider that $f(\mathbf{p},t)$ is a series with two indices $\big(n,m\big)$. So, the distribution function is expanded in terms of $\phi$ and $t$ harmonics with coefficients $f_m^{(n)}$ that are functions of $\epsilon$, the total energy of an electron
\begin{equation} \label{fmn}
f(\mathbf{p},t)=\sum_{n,m}{f^{(n)}_m \exp(im\phi-in\omega t)},\nonumber
\end{equation}
where $m$ and $n$ are integers and multiplying the Boltzmann equation by a factor
$ \exp { ( - i j \phi + i l \omega  t) }$,
where $j$ and $l$ are integers and integrating over a period $2\pi$ of angle $\phi$ and a period of time, $t$, lead to coupled equations between different harmonic coefficients
\begin{eqnarray}\label{fjl}
(\tau _{\left | j \right |,\mathbf{p}}^{-1}-il\omega)f_{j}^{l}&&=\alpha _{j-1}f_{j-1}^{l-1}+\widetilde{\alpha}_{j-1}f_{j-1}^{l+1}+\eta _{j+1}f_{j+1}^{l-1}\nonumber\\
&&+\widetilde{\eta}_{j+1}f_{j+1}^{l+1}+\delta S_{j}^{l}.\nonumber\\
\end{eqnarray}
For an isotropic material in the absence of the magnetic field
\begin{eqnarray}
\tau^{-1}_{\left | j \right |}\equiv \sum_{p^{\prime}}W_{p^{\prime}p}[1-\cos(j[\phi^{\prime}-\phi])]\nonumber
\end{eqnarray}
is the relaxation time of the $jth$ angular harmonic of the electron distribution function. However, for an anisotropic 2DEG like phosphorene, it is \cite{relaxtime, lowrelax} 
\begin{eqnarray}
\tau^{-1}_{\left | j \right |, \mathbf {p}}(\mathbf \xi, \mathbf p )&&=\frac{2\pi}{\hbar}\sum_{\mathbf p^{\prime}}  \left| \langle \mathbf{p^{\prime}}\left| \delta H \right|\mathbf{p}\rangle\right|^2 \delta \left(\epsilon_p-\epsilon_{p^{\prime}} \right)\nonumber\\
&& \quad \times \left\{1-\frac{ [ \mathbf{\xi} . \mathbf{V}_g(\mathbf{p^{\prime}}) ] \tau_{\left | j \right |, \mathbf {p^{\prime}}}}{[ \mathbf{\xi} . \mathbf {V}_g(\mathbf{p}) ] \tau_{\left | j \right |, \mathbf {p}}} \right\},\nonumber
\end{eqnarray}  
where $\xi$ is the unit matrix of the electric field and $\mathbf {V}_g$ is the group velocity.   
In addition, operators in Eq.~\ref{fjl} are 
\begin{equation}
\alpha_j= \frac{e (E_x-iE_y)}{2} \left(-\frac{j}{p} +\frac{\partial }{\partial p} \right ),\nonumber
\end{equation}
\begin{equation}
\tilde{\alpha}_j= \frac{e (E_{x}^{*}-iE_{y}^{*})}{2} \left(-\frac{j}{p} +\frac{\partial }{\partial p} \right ),\nonumber
\end{equation}
\begin{equation}
\eta_j= \frac{e (E_x+iE_y)}{2} \left (\frac{j}{p} +\frac{\partial }{\partial p}\right ),\nonumber
\end{equation}
\begin{equation}
\tilde{\eta}_j= \frac{e (E_{x}^{*}+iE_{y}^{*})}{2} \left( \frac{j}{p} +\frac{\partial }{\partial p} \right ),\nonumber
\end{equation}
where $p=\left|\mathbf{p}\right|$. The factors $\delta S_{j}^{l}$ in Eq.~\ref{fjl} describe the correction to the scattering caused by the magnetic field. 

To quantify the dc current caused by an ac electric field, it is necessary to determine time--independent asymmetric parts of the distribution function; $f^{0}_{\pm1}$ harmonics.
We assume that electrons are trapped in a huge box with length $L$ and under a periodic potential. For $
\delta f=f_{1}^{0}\exp(i\phi)+f_{-1}^{0}\exp(-i\phi)$, the current density is
\begin{eqnarray}\label{current}
\mathbf{J}=-\frac{g}{L^2}\sum_{\mathbf{p}}e \mathbf{V}_{g} \delta f,
\end{eqnarray}
where $g$ is the spin degeneracy factor ($g=2$). 

To solve Eq.~\ref{current}, the summation over the momentum vector could be written based on two integrals, one integral over the electron energy, $\epsilon$, and another one over the direction of the momentum, $\phi$
\begin{equation}
\sum_\mathbf{p}(...)=L^2\int_{0}^{\infty}\Gamma(\epsilon)d\epsilon\int_{0}^{2 \pi}\frac{d\phi}{2\pi}(...).\nonumber
\end{equation}
The coupled equations (Eq.~\ref{fjl}) for the harmonics $f_j^{(l)}$ of the distribution function are also used to express $\delta f$ in terms of the equilibrium distribution function $f_0^{(0)}$. In addition, we consider that we have a degenerate electron gas, at low--temperature condition; $k_B T \ll \epsilon _f$. Hence, we can assume that $\partial f_0^{(0)}/ \partial \epsilon \approx-\delta(\epsilon-\epsilon_f)$. 
\section{Phosphorene}
We assume that the band dispersion is equal to $\epsilon$, density of states per spin per unit area is $\Gamma (\epsilon)$, and the group velocity of trapped electrons is  $\mathbf{V}_{g}=\nabla_{\mathbf{p}} \epsilon$. 
To calculate the change of the scattering rate, $\delta W_{p^{\prime} p}$ in Eq.~\ref{sr}, we use a numerical approach. Accordingly, after deriving the perturbed eigenstates of the Hamiltonian (Eq.~\ref{HBP1}) according to Eq.~\ref{perteigen}, we substitute values of $t_i$ $(i=1, 2 ... 5)$ coupling parameters and lattice parameters in derived perturbed eigenstate for the conduction and valence bands. We consider that $t_1 =-1.220 eV$, $t_2 = 3.665 eV$, $t_3 =-0.205 eV$, $t_4 =-0.105 eV$ and $t_5 =-0.055 eV$ \cite{ezawa,chaves2017theoretical, pereira2015landau}. We also substitute numerical values of $a_x$, $a_y$, $\alpha$, $\beta$ and $d^{\prime}$ (Eqs.~\ref{f1} to \ref{f5}). To derive $W_{\mathbf{p^\prime} \mathbf{p}}$, we also consider different forms of $\hat{Y}$ matrix (Eqs.~\ref{formulsp} and \ref{deltah}). In the symmetric case, where the upper and lower layers are under the effect of disorder, with equal amounts of scattering on the two layers, the disorder matrix is a unit matrix. Besides, according to the selected basis that is $(A_1,B_1,A_2,B_2)^{T}$, if we consider that scattering is limited to the lower layer ($\zeta=1$) or upper layer ($\zeta=-1$), the disorder matrix is 
\begin{eqnarray}
\hat{Y}=\frac {1}{2} \big( \hat{I}+\zeta \hat{\sigma_z}\otimes\hat{I} \big).\nonumber
\end{eqnarray}
Hence, we derive a general form for $W_{\mathbf{p^{\prime}} \mathbf{p}}$. This general form that is linear in magnetic field and momentum is
\begin{eqnarray}\label{deltaw}
W_{\mathbf{p^{\prime}}\mathbf{p}} (U_1,U_2,\delta,\zeta) && = \frac{2 \pi}{\hbar} \frac{n_{imp}}{L^2}  \mid \tilde{u}( \bf{p}^{\prime}- \bf{p} ) \mid ^ 2 \delta(\epsilon_{p^\prime}-\epsilon_p)\nonumber\\
\quad && \times \bigg\{C_0 (U_1,U_2,\delta, \zeta)\nonumber\\
\quad && + \frac{1}{\hbar^2} C_1(U_1,U_2,\delta, \zeta) b_y p (\cos \phi + \cos\phi^ \prime)\nonumber\\
&& + \frac{1}{\hbar^2} C_2 (U_1,U_2,\delta,\zeta) b_x p ( \sin \phi +\sin \phi^ \prime ) \bigg\},\nonumber 
\end{eqnarray}
where $n_{imp}=N_{imp}/L^2$ is the density of impurities and $C_0$, $C_1$ and $C_2$ are three parameters that change by the change of on--site energies ($U_1$, $U_2$ and $\delta$) and disorder types. To estimate $C_0$, $C_1$ and $C_2$ prefactors, we discuss about the problem numerically. We select $U_1=0$, and we consider different values for $U_2$ and $\delta$ parameters. Hence, $U_2$ is a tunable factor that is defined as the difference in the electrostatic potential on the two layers \cite{edratchet, novo}. We consider $\delta$ in the range of 0 to $20 meV$ and $U_2$ in the range of $0$ to $40 m eV$ \cite{edratchet, novo}. Then, we calculate $C_0$, $C_1$ and $C_2$ based on selected ranges of values for three different disorder types. 

For the case of symmetric disorder, $\hat{Y}=\hat{I}$, we can show that the ratchet current is equal to zero; $C_1=C_2=0$. Consequently, the symmetry of the upper and lower layer should be broken by disorder or substrate to have a nonzero ratchet current.

For asymmetric disorder types, we can show that by the change of $\delta$ in the range of 0 to $20 meV$ and $U_2$ in the range of $0$ to $40 m eV$, the $C_0$ is equal to $1/4$ in order of $10^{-2}$. Additionally, for these two disorder types, for the conduction and valence bands, we can show that considering $\delta=U_2=0$ will deduce to a zero ratchet current. Besides, for any amount of $\delta$ factor, for a nonzero $U_2$ amount, $C_1$ is nonzero. However for $\delta=0$ or $U_2=0$, $C_2$ is equal to zero.  

Moreover, we can show that the relevant $\delta S_{j}^{l}$ factors in Eq.~\ref{fjl} that describe the correction to scattering caused by the magnetic field are  
\begin{allowdisplaybreaks}
\begin{eqnarray}
\delta S_{0}^{l}&&=0, \nonumber\\
\delta S_{1}^{l}&&=\Lambda  \big( C_1 B_y +i C_2 B_x\big) f_2^l,\nonumber\\
\delta S_{-1}^{l}&&=\Lambda  \big( C_1 B_y -i C_2 B_x \big) f_{-2}^l,\nonumber\\
\delta S_{2}^{l}&&=\Lambda  \big( C_1 B_y - i C_2 B_x\big) f_1^l,\nonumber\\
\delta S_{-2}^{l}&&=\Lambda  \big(C_1 B_y + i C_2 B_x\big) f_{-1}^l,\nonumber
\end{eqnarray}
\end{allowdisplaybreaks}
where 
\begin{eqnarray}
\Lambda && = \frac{e d \pi n_{imp}}{2 \hbar^3} \Omega  \Gamma(\epsilon) p ,\nonumber\\
\Omega &&= -(\nu_0-\nu_2).\nonumber
\end{eqnarray}
Hence, we can show that the corresponding in--plane current is
\begin{eqnarray}\label{jx11}
J_x&&=M_{1,x}[B_y^\prime(\left | E_x \right |^2-\left | E_y \right |^2) + B_x^\prime(E_xE_y^*+E_yE_x^*)]\nonumber\\
&&+M_{2,x}B_y^\prime\left | E \right |^2 +M_{3,x} B_x^\prime i(E_xE_y^*-E_yE_x^*),
\end{eqnarray}
\begin{eqnarray}\label{jy11}
J_y&&=M_{1,y}[- B_x^\prime(\left | E_x \right |^2-\left | E_y \right |^2)+B_y^\prime(E_xE_y^*+E_yE_x^*)]\nonumber\\
&& + M_{2,y} B_x^\prime\left | E \right |^2 - M_{3,y} B_y^\prime i(E_xE_y^*-E_yE_x^*), 
\end{eqnarray}
where $B_y^\prime= C_1 B_y$ and $B_x^\prime= C_2 B_x$. Furthermore, $M$ coefficients are the current responses to different light polarizations. $M_1$ is the current response to the linearly polarized light, $M_2$ is the current response to the unpolarized light, and $M_3$ is the current response to the circularly polarized light. We can show that for phosphorene and anisotropic 2D materials
\begin{eqnarray}
M_{1,i}=-\frac{ g e^3}{4 L^2}\sum_\mathbf{p} V_{g,i} \tau_{1,i} \tau_{2,i} \Lambda \bigg(\frac{-1}{p}+\frac{\partial }{\partial p} \bigg) 
 \frac{2 \tau_{1,i}}{1 + \omega^2 \tau_{1,i}^2}\frac{\partial f_0}{\partial p}, \nonumber
\end{eqnarray}
\begin{eqnarray}
M_{2,i}&&=-\frac{ g e^3}{4 L^2}\sum_\mathbf{p} V_{g,i} \tau_{1,i} \bigg(\frac{2 }{p}+\frac{\partial }{\partial p} \bigg)\nonumber\\
&&\quad \times \frac{\big( 2\tau_{1,i} \tau_{2,i} \Lambda \big)\big(1-\omega^2 \tau_{1,i} \tau_{2,i} \big)}{\big(1+\omega^2 \tau_{1,i}^2 \big)\big(1+\omega^2 \tau_{2,i}^2\big)}\frac{\partial f_0}{\partial p},\nonumber
\end{eqnarray}
\begin{eqnarray}
M_{3,i}&&=-\frac{ g e^3}{4 L^2}\sum_\mathbf{p} V_{g,i} \tau_{1,i} \bigg(\frac{2 }{p}+\frac{\partial }{\partial p} \bigg)\nonumber\\ 
 && \quad \times \frac {\big( 2 \omega \tau_{1,i} \tau_{2,i} \Lambda \big)\big(\tau_{1,i}+\tau_{2,i})}{\big(1+\omega^2 \tau_{1,i}^2 \big)\big(1+\omega^2 \tau_{2,i}^2\big)}\frac{\partial f_0}{\partial p},\nonumber
\end{eqnarray}
where $V_{g,x}=p/m_{xx}$, $V_{g,y}=p/m_{yy}$, and $m_{xx}$ and $m_{yy}$ are the effective masses along $x$ (armchair) and $y$ (zigzag) directions.
\section{Discussion}
To determine $M$ coefficients and the current, we assume that the eigenvalues of the system are based on the Ref.~\cite{AsgariHamil}. Accordingly, 
\begin{eqnarray}
\frac{\partial }{\partial p}=C_{ph} p \frac{\partial }{\partial \epsilon},\nonumber
\end{eqnarray} 
and
\begin{eqnarray}
C_{ph}=s\frac{2}{\hbar^2}\bigg[ \frac{\gamma^2}{E_g}+\big(\eta_{v/c}+\nu_{v/c}\big)\bigg],\nonumber
\end{eqnarray}
where $s$ is the band index and it is $+1$ for the conduction band and $-1$ for the valence band. $E_g$ is the direct energy gap, $\gamma=0.480eV nm^2$, $\eta_v=0.038 eV nm^{2}$, $\nu_v=0.030 eV nm^2$, $\eta_c= 0.008 eV nm^{2}$ and $\nu_c=0.030 eV nm^2$ are from Ref.~\cite{AsgariHamil}. The above values have been calculated  for $E_g=0.912 eV$ that it can be potentially tuned \cite{AsgariHamil,eg}. Hence, we can show that 
\begin{eqnarray}\label{m1}
M_{1,i}&&=-\frac{g e^3}{2}C_{ph}\frac{\tau_{1,i}}{1+ \tau^2_{1,i}\omega^2}\ \nonumber\\
&&\times\bigg[V_{g,i}\Gamma(\epsilon)\tau_{1,i}\tau_{2,i}\Lambda+C_{ph}p \big(  \Gamma(\epsilon) V_{g,i} \tau_{1,i} \tau_{2,i} \Lambda p \big)^{\prime} \bigg], \nonumber\\
\end{eqnarray}
\begin{eqnarray}\label{m2}
M_{2,i}&&=\frac{g e^3}{2}C_{ph}\frac{\tau_{1,i}\tau_{2,i} \Lambda (1-\omega^2 \tau_{1,i} \tau_{2,i})}{(1+\tau^2_{1,i}\omega^2)(1+\tau^2_{2,i}\omega^2)}\ \nonumber\\
&& \times \bigg[2 V_{g,i}\Gamma(\epsilon)\tau_{1,i}- C_{ph}p \big(  \Gamma(\epsilon) V_{g,i} \tau_{1,i} p \big)^{\prime} \bigg],
\end{eqnarray}
\begin{eqnarray}\label{m3}
M_{3,i}&&=\frac{g e^3}{2}C_{ph}\frac{\omega \tau_{1,i}\tau_{2,i} \Lambda (\tau_{1,i}+ \tau_{2,i})}{(1+\tau^2_{1,i}\omega^2)(1+\tau^2_{2,i}\omega^2)}\ \nonumber\\ 
&& \times \bigg[2 V_{g,i}\Gamma(\epsilon)\tau_{1,i}-C_{ph}p \big(  \Gamma(\epsilon) V_{g,i} \tau_{1,i} p \big)^{\prime} \bigg].
\end{eqnarray}
Here, derivatives are related to the energy, $(...)^\prime\equiv \partial (...)/\partial \epsilon$, and all parameters are evaluated on the Fermi surface. We have also assumed that $E_f=\hbar^2 \pi n /m_d$; $n$ is the carrier density in phosphorene \cite{lowrelax} and $m_d=\sqrt{m_{xx} m_{yy}}$ where $m_{xx}=0.15 m_0$, $m_{yy}=0.7 m_0$ and $m_0$ is electron free mass \cite{lowrelax}. In addition, we have $V_{g,x}=2 s p(\gamma^2+E_g \eta_{v/c})/E_g \hbar^2$ and $V_{g,y}=2 s p \nu_{v/c} / \hbar^2$. We also consider the density of states, $\Gamma (\epsilon)$, equal to $m_d/\pi \hbar^2$. As we discussed before, the momentum relaxation time is dependent on the electric field direction as well as the incoming wave vector \cite{lowrelax}. Consequently, the current responses to different light polarizations, $M$ coefficients, are not equal to zero. Hence, phosphorene and anisotropic 2D materials have responses to three types of radiation: the unpolarized light, the linearly polarized light and the circularly polarized light.

For linearly polarized light, we can assume that $E_x^{*}=E_x=(E_0/2) \cos{\theta}$ and $E_y^{*}=E_y=(E_0/2) \sin{\theta}$, where $\theta$ is the polarization angle. We also assume that $B^{\prime}_\parallel=\big({B^{\prime}_x}^2+{B^{\prime}_y}^2\big)^{1/2}$ and $\varphi^{\prime}=\arctan\big( B^{\prime}_y / B^{\prime}_x \big)$. Consequently, we can show that the current density is 
\begin{eqnarray}
J_x&&=\frac{E_0^2}{4}B^{\prime}_\parallel \big( M_{1,x} \sin({2 \theta+\varphi^{\prime}})+M_{2,x} \sin{\varphi^{\prime}}\big),\nonumber\\
J_y&&=\frac{E_0^2}{4}B^{\prime}_\parallel \big( -M_{1,y} \cos({2 \theta+\varphi^{\prime}})+M_{2,y} \cos{\varphi^{\prime}}\big).\nonumber
\end{eqnarray}  
According to above equations, the direction of the current density for the linearly polarized light is dependent on the polarization angle $\theta$, the magnetic field direction, the gate voltage and the disorder location what determine $\varphi^{\prime}$.
Also, for unpolarized light, $M_{1,x/y}$ related terms are equal to zero while the $M_{2,x/y}$ related terms survive.
For circularly polarized light, we can show that $E_x^{*}=E_x=E_0/2$ and $E_y^{*}=-E_y=- i \mu E_0/2$, where $\mu=1(-1)$ for the left- (right-) handed circularly polarized light. For circularly polarized light, we can show that 
\begin{eqnarray}
J_x&&=\frac{E_0^2}{2}B^{\prime}_\parallel \big( M_{2,x} \sin{\varphi^{\prime}}+\mu M_{3,x} \cos{\varphi^{\prime}}\big),\nonumber\\
J_y&&=\frac{E_0^2}{2}B^{\prime}_\parallel \big( M_{2,y} \cos{\varphi^{\prime}}+\mu M_{3,y} \sin{\varphi^{\prime}}\big).\nonumber
\end{eqnarray}
For the circularly polarized light, the current density direction is determined by the $\mu$, the magnetic field direction, the gate voltage and the disorder location.
 
Furthermore, based on the momentum relaxation time and the group velocity, the current magnitude changes. Besides, dependent on the place of the disorder, the effect of an applied magnetic field in $x$ and $y$ direction changes. So, the macroscopic current has the sign of the microscopic occurrence. Note that the frequency dependencies of $M$ coefficients for the isotropic and anisotropic materials are similar \cite{edratchet}.

To estimate the strength of the effect, we use parameters of Ref.~\cite{lowrelax} and we consider that momentum relaxation time is independent of the energy and $\zeta$ is equal to 1. Hence, we assume for carrier densities $10^{16} m ^{-2}$, $n_{imp}=10^{16} m^{-2}$, and for impurity distance $0 n m$, we have $\tau_{x} \approx \tau_{y}=0.1 ps$ \cite{lowrelax}. We also assume that $\nu_0$ is independent of the energy, and it is equal to what have been calculated for bilayer graphene \cite{edratchet}, $p_f$ is of order of $10^{-26}$ $kg. m s ^{-1}$, $|E| = 10 kV cm^{-1}$, $|B| = 7 T$, and $\omega = 2.1 \times 10^{13} rad \times s^{-1}$ \cite{taranature}. In the case of valence band, for $\zeta=1$, $\delta= 0.02 eV$ and $U_2= 0.04 eV$, $C_1=0.034 \AA ^{2}$ and $C_2=-1.882 \times 10^{-8} \AA ^{2}$. It is worthy to mention that, for the conduction band, the magnitude of $C_1$ and $C_2$ prefactors are similar to the valence band. For instance, in the case of conduction band where $\delta= 0.02 eV$, and $0<U_2<0.04$, $C_1$ prefactor decreases linearly between $0$ and $ -0.034 \AA ^2$, and $C_2$ prefactor increase linearly from $0$ to $1.861 \times 10^{-8} \AA ^2$. 
Hence, according to Eqs.~\ref{jx11} to \ref{m3}, we estimate that the magnetic ratchet current density caused by the applied linearly and circularly polarized lights for $x$ and $y$ directions are in order of $\mu A cm^{-1}$. And, for unpolarized light the current density for $x$ direction is in order of $\mu A cm^{-1}$ and for $y$ direction is in order of $n A cm^{-1}$.  
\section{Conclusion}
We have considered phosphorene to study the ratchet current in the anisotropic materials. The tight--binding Hamiltonian of phosphorene in a parallel magnetic field has been derived and the orbital effect of an in--plane magnetic field on electrons in phosphorene is demonstrated. Moreover, the semi--classical Boltzmann kinetic equation is used to derive the direct current in phosphorene under the in--plane magnetic field. For anisotropic materials under an asymmetric disorder or substrate, ratchet current includes the response to three types of radiations: linearly polarized light, circularly polarized light, and unpolarized light. This result can play an important role in inorganic material based polarized and unpolarized light detectors.
\bibliography{bib} 

\begin{thebibliography}{10}

\bibitem{tarasenko2008electron}
S.~Tarasenko, ``Electron scattering in quantum wells subjected to an in-plane
  magnetic field,'' {\em Physical Review B}, vol.~77, no.~8, p.~085328, 2008.

\bibitem{budkin2016ratchet}
G.~Budkin and S.~Tarasenko, ``Ratchet transport of a two-dimensional electron
  gas at cyclotron resonance,'' {\em Physical Review B}, vol.~93, no.~7,
  p.~075306, 2016.

\bibitem{taranature}
C.~Drexler, S.~Tarasenko, P.~Olbrich, J.~Karch, M.~Hirmer, F.~M{\"u}ller,
  M.~Gmitra, J.~Fabian, R.~Yakimova, S.~Lara-Avila, {\em et~al.}, ``Magnetic
  quantum ratchet effect in graphene,'' {\em Nature nanotechnology}, vol.~8,
  no.~2, pp.~104--107, 2013.

\bibitem{PhysRevB.93.075422}
P.~Olbrich, J.~Kamann, M.~K\"onig, J.~Munzert, L.~Tutsch, J.~Eroms, D.~Weiss,
  M.-H. Liu, L.~E. Golub, E.~L. Ivchenko, V.~V. Popov, D.~V. Fateev, K.~V.
  Mashinsky, F.~Fromm, T.~Seyller, and S.~D. Ganichev, ``Terahertz ratchet
  effects in graphene with a lateral superlattice,'' {\em Phys. Rev. B},
  vol.~93, p.~075422, Feb 2016.

\bibitem{simosfet}
S.~D. Ganichev, S.~A. Tarasenko, J.~Karch, J.~Kamann, and Z.~D. Kvon,
  ``Magnetic quantum ratchet effect in si-mosfets,'' {\em Journal of Physics:
  Condensed Matter}, vol.~26, no.~25, p.~255802, 2014.

\bibitem{edratchet}
N.~Kheirabadi, E.~McCann, and V.~I. Fal'ko, ``Magnetic ratchet effect in
  bilayer graphene,'' {\em Physical Review B}, vol.~94, no.~16, p.~165404,
  2016.

\bibitem{kheirabadi2018cyclotron}
N.~Kheirabadi, E.~McCann, and V.~I. Fal'ko, ``Cyclotron resonance of the
  magnetic ratchet effect and second harmonic generation in bilayer graphene,''
  {\em Physical Review B}, vol.~97, no.~7, p.~075415, 2018.

\bibitem{direct}
S.~Tarasenko, ``Direct current driven by ac electric field in quantum wells,''
  {\em Physical Review B}, vol.~83, no.~3, p.~035313, 2011.

\bibitem{viti2015black}
L.~Viti, J.~Hu, D.~Coquillat, W.~Knap, A.~Tredicucci, A.~Politano, and M.~S.
  Vitiello, ``Black phosphorus terahertz photodetectors,'' {\em Advanced
  Materials}, vol.~27, no.~37, pp.~5567--5572, 2015.

\bibitem{sun2016optical}
Z.~Sun, A.~Martinez, and F.~Wang, ``Optical modulators with 2d layered
  materials,'' {\em Nature Photonics}, vol.~10, no.~4, pp.~227--238, 2016.

\bibitem{lee2019fabrication}
Y.~Lee, S.~Lee, J.-Y. Yoon, J.~Cheon, H.~Y. Jeong, and K.~Kim, ``Fabrication
  and imaging of monolayer phosphorene with preferred edge configurations via
  graphene-assisted layer-by-layer thinning,'' {\em Nano letters}, vol.~20,
  no.~1, pp.~559--566, 2019.

\bibitem{akhtar2017recent}
M.~Akhtar, G.~Anderson, R.~Zhao, A.~Alruqi, J.~E. Mroczkowska, G.~Sumanasekera,
  and J.~B. Jasinski, ``Recent advances in synthesis, properties, and
  applications of phosphorene,'' {\em npj 2D Materials and Applications},
  vol.~1, no.~1, pp.~1--13, 2017.

\bibitem{carvalho2016phosphorene}
A.~Carvalho, M.~Wang, X.~Zhu, A.~S. Rodin, H.~Su, and A.~H.~C. Neto,
  ``Phosphorene: from theory to applications,'' {\em Nature Reviews Materials},
  vol.~1, no.~11, pp.~1--16, 2016.

\bibitem{falko1989rectifying}
V.~Falko, ``Rectifying properties of 2d inversionlayers in a parallel
  magnetic-field,'' {\em Fizika Tverdogo Tela}, vol.~31, no.~4, pp.~29--32,
  1989.

\bibitem{chen2019circularly}
C.~Chen, L.~Gao, W.~Gao, C.~Ge, X.~Du, Z.~Li, Y.~Yang, G.~Niu, and J.~Tang,
  ``Circularly polarized light detection using chiral hybrid perovskite,'' {\em
  Nature communications}, vol.~10, no.~1, pp.~1--7, 2019.

\bibitem{novo}
A.~Kuzmenko, I.~Crassee, D.~Van Der~Marel, P.~Blake, and K.~Novoselov,
  ``Determination of the gate-tunable band gap and tight-binding parameters in
  bilayer graphene using infrared spectroscopy,'' {\em Physical Review B},
  vol.~80, no.~16, p.~165406, 2009.

\bibitem{chaves2017theoretical}
A.~Chaves, W.~Ji, J.~Maassen, T.~Dumitrica, and T.~Low, ``Theoretical overview
  of black phosphorus,'' in {\em 2D Materials: Properties and Devices},
  pp.~381--412, Cambridge University Press, 2017.

\bibitem{pereira2015landau}
J.~Pereira~Jr and M.~Katsnelson, ``Landau levels of single-layer and bilayer
  phosphorene,'' {\em Physical Review B}, vol.~92, no.~7, p.~075437, 2015.

\bibitem{glazov2014high}
M.~Glazov and S.~Ganichev, ``High frequency electric field induced nonlinear
  effects in graphene,'' {\em Physics Reports}, vol.~535, no.~3, pp.~101--138,
  2014.

\bibitem{relaxtime}
Y.~Liu and P.~P. Ruden, ``Temperature-dependent anisotropic charge-carrier
  mobility limited by ionized impurity scattering in thin-layer black
  phosphorus,'' {\em Physical Review B}, vol.~95, no.~16, p.~165446, 2017.

\bibitem{lowrelax}
Y.~Liu, T.~Low, and P.~P. Ruden, ``Mobility anisotropy in monolayer black
  phosphorus due to scattering by charged impurities,'' {\em Physical Review
  B}, vol.~93, no.~16, p.~165402, 2016.

\bibitem{ezawa}
M.~Ezawa, ``Topological origin of quasi-flat edge band in phosphorene,'' {\em
  New Journal of Physics}, vol.~16, no.~11, p.~115004, 2014.

\bibitem{AsgariHamil}
M.~Zare, B.~Z. Rameshti, F.~G. Ghamsari, and R.~Asgari, ``Thermoelectric
  transport in monolayer phosphorene,'' {\em Physical Review B}, vol.~95,
  no.~4, p.~045422, 2017.

\bibitem{eg}
R.~Fei, A.~Faghaninia, R.~Soklaski, J.-A. Yan, C.~Lo, and L.~Yang, ``Enhanced
  thermoelectric efficiency via orthogonal electrical and thermal conductances
  in phosphorene,'' {\em Nano letters}, vol.~14, no.~11, pp.~6393--6399, 2014.

\end{thebibliography}
\bibliographystyle{ieeetr}

\end{document}